\documentclass[twocolumn]{aastex63}
\let\tablenum\relax
\usepackage[T1]{fontenc}
\usepackage{multirow}
\usepackage{natbib}
\usepackage[version=4]{mhchem}
\usepackage{chemmacros}

\chemsetup{
    modules = all,
    formula = mhchem
    }

\graphicspath{figures/}

\shortauthors{Canta et al.}

\begin{document}

\title{The First Detection of CH$_2$CN in a Protoplanetary Disk}

\author[0000-0002-7588-8054]{Alessandra Canta}
\affil{Center for Astrophysics | Harvard \& Smithsonian,
       60 Garden Street, 
       Cambridge,
       MA 02138,
       USA}
       
\author[0000-0002-0786-7307]{Richard Teague}
\affil{Center for Astrophysics | Harvard \& Smithsonian,
       60 Garden Street, 
       Cambridge,
       MA 02138,
       USA}

\author[0000-0003-1837-3772]{Romane Le Gal}
\affil{Center for Astrophysics | Harvard \& Smithsonian,
       60 Garden Street, 
       Cambridge,
       MA 02138,
       USA}
\affiliation{Univ. Grenoble Alpes, CNRS, IPAG, F-38000 Grenoble, France}
\affiliation{IRAM, 300 rue de la piscine, F-38406 Saint-Martin d'H\`{e}res, France}
       
\author[0000-0001-8798-1347]{Karin I. Öberg}
\affil{Center for Astrophysics | Harvard \& Smithsonian,
       60 Garden Street, 
       Cambridge,
       MA 02138,
       USA}

\begin{abstract}
\noindent We report the first detection of the molecule cyanomethyl, \ce{CH2CN}, in a protoplanetary disk. Until now, \ce{CH2CN} had only been observed at earlier evolutionary stages, in the giant molecular clouds TMC-1 and Sgr 2, and the prestellar core L1544. We detect six transitions of ortho-\ce{CH2CN} towards the disk around nearby T Tauri star TW Hya. An excitation analysis reveals that the disk-averaged column density, $N$, for ortho-\ce{CH2CN} is $(6.3\pm 0.5)\times10^{12}$~cm$^{-2}$, which is rescaled to reflect a 3:1 ortho-para ratio, resulting in a total column density, $N_{\rm tot}$, of $(8.4\pm 0.7)\times10^{12}$~{\rm cm}$^{-2}$. We calculate a disk-average rotational temperature, $T_{\rm{rot}}$ = $40 \pm 5$~K, while a radially resolved analysis shows that $T_{\rm{rot}}$ remains relatively constant across the radius of the disk. This high rotation temperature suggests that in a static disk and if vertical mixing can be neglected, \ce{CH2CN} is largely formed through gas-phase reactions in the upper layers of the disk, rather than solid-state reactions on the surface of grains in the disk midplane. The integrated intensity radial profiles show a ring structure consistent with molecules such as CN and DCN. We note that this is also consistent with previous lower-resolution observations of centrally peaked CH$_3$CN emission towards the TW Hya disks, since the observed emission gap disappears when convolving our observations with a larger beam size. We obtain a \ce{CH2CN}/\ce{CH3CN} ratio ranging between 4 and 10. This high \ce{CH2CN}/\ce{CH3CN} is reproduced in a representative chemical model of the TW Hya disk that employs standard static disk chemistry model assumptions, i.e. without any additional tuning. 
\end{abstract}


\section{Introduction}

\begin{figure*}[hpt!]
\centering
\includegraphics[width=0.9\textwidth]{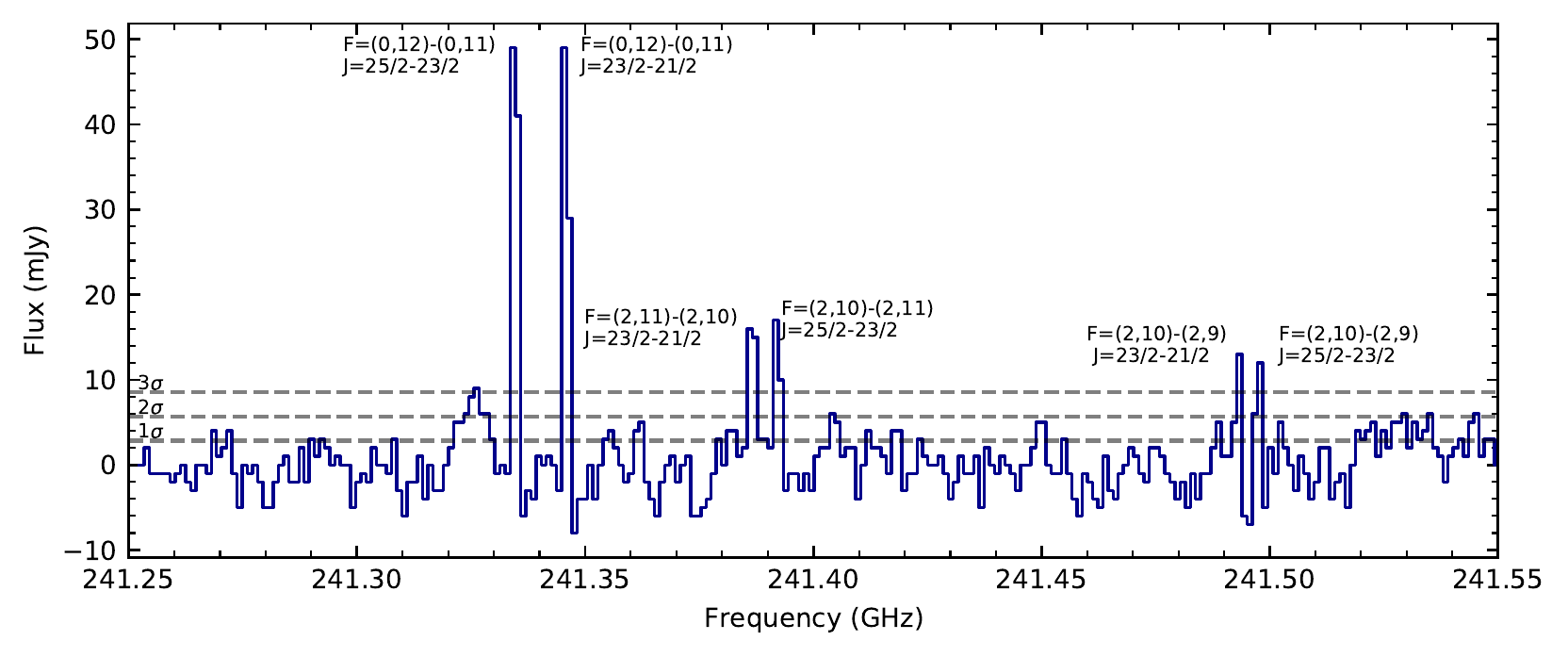}
\caption{Spectrum of the observed \ce{CH2CN} transitions. The 6 brightest peaks are shown with their quantum numbers. The noise is calculated to be 2.9 mJy, with the horizontal dashed lines showing 1, 2 and $3~\sigma$ levels.}
\label{fig:frequency_spectrum}
\end{figure*}

The composition of planets, and therefore their suitability to host biological life, is largely dictated by the physical structure and chemical composition of ice and dust grains in the protoplanetary disks surrounding young stars \citep[e.g.,][]{Nomura_2016, Andrews_2012}. Nitriles are of specific interests since they have, in fact, often been identified as key precursors in the synthesis of RNA and amino acids \citep{Powner_2009, Patel_2015}. Observations of comets and meteors reveal that molecules such as HCN and \ce{CH3CN} were present in the chemical environment of our early Solar Nebula, and were likely available for prebiotic reactions \citep{Mumma_2011, Altwegg_2020}. Nitrile existence in both our early Solar system and in the disks of young stars has implications for our understanding of the trajectory of the organic chemistry as it evolves during planet formation. This is important to estimate the organic inventory on newly-formed planets and to determine what factors contribute to the chemical habitability of nascent planets \citep{Bergner_2018}.

HCN and CN were among the first compounds to be observed in a protoplanetary disk, and since their detection in 1997 over 25 new species have been identified in disks \citep{Kastner_1997, McGuire_2018}. However, it was not until the advent of ALMA that more complex molecules such as acetonitrile, \ce{CH3CN}, and methanol, \ce{CH3OH}, were observed \citep{Oberg_2015, Walsh_2016}. Other nitriles that have been detected include \ce{HNC} and cyanopolyyne, \ce{HC3N}, as well as various isotopologues of HCN and CN, including H$^{13}$CN, HC$^{15}$N, DCN and C$^{15}$N \citep{Dutrey_1997, Chapillon_2012, Guzman_2015, Qi_2003, Hily-Blant_2017}. Together these nitriles constitute a substantial fraction of the detected organics in disks, which points to an interesting variations in chemical composition at different stages of a star's life: the early stellar stages seem to have significantly less oxygen-rich species than protoplantary disks \citep{Oberg_2020}. This could be as a result of the oxygen being locked up in molecules such as CO and \ce{H2O} and thus unavailable for chemistry, or of a generally oxygen-poor grain-surface chemistry \citep{Oberg_2020}. 

The formation of complex organic molecules (COMs) in protoplanetary disks can happen via two distinct pathways: through grain-surface or gas-phase reactions. Freeze-out reactions on the surface of grains seem to be the greatest contributors to the abundance of COMs in disks, and they involve the absorption of UV radiation from both the central star and the interstellar radiation field (ISRF), formation of radicals and subsequent rearrangement into more complex organics \citep{Oberg_2016}. Gas-phase reactions are dominated by a variety of reactions including radiative association (molecules collide and emit a photon), neutral-neutral and neutral-ion reactions and dissociative recombination, where a positive ion recombines with an electron. This was found to be the main reaction pathway for the formation of \ce{CH3CN} in TW~Hya and of \ce{CH2CN} in the protostellar core L1544 \citep{Loomis_2018, Vastel_2015}. Whether this is also the case for \ce{CH2CN} in disks has not previously been possible to evaluate since there have been no reported observations of this molecule.

In this paper, we report the first observation of the molecule \ce{CH2CN} (in its ortho state) in a protoplanetary disk, specifically in the disk around the nearby T Tauri star TW Hya. This molecule has previously been observed in diffuse molecular clouds SgrB2 and TMC-1, in the pre-stellar core L1544, and in the circumstellar envelope surrounding the star IRC~+10216 but never in a protoplanetary disk \citep{Vastel_2015, Irvine_1988, Agundez_2008}. \citet{Liszt_2018} also attempted to place upper limits on the abundance of \ce{CH2CN} in diffuse molecular gas. TW Hya is a $\sim0.8M_\odot$ solar-like T Tauri star that is often used in astrochemical observations because of its proximity  \citep[60.1~pc][]{Bailer-Jones_ea_2018} and its face-on orientation, $i \approx 5\degr$, which allows for easier interpretation of the data. The details of the detection, the data reduction and the observational results are outlined in \S \ref{section: 2}. In \S \ref{section: excitation_analysis}, we proceed by using a disk-averaged rotational diagram analysis to obtain the total column density for ortho-\ce{CH2CN} and an excitation temperature. We also perform a radially resolved analysis to obtain radial profiles for both these parameters. In \S \ref{section: discussion} we compare our findings to a chemical model of TW~Hya, which we use to complement our discussion of the chemistry of both \ce{CH3CN} and \ce{CH2CN}. Finally, we present a summary of our findings in \S \ref{section: summary}.

\begin{figure*}[hpt!]
\centering
\includegraphics[width=0.8\textwidth]{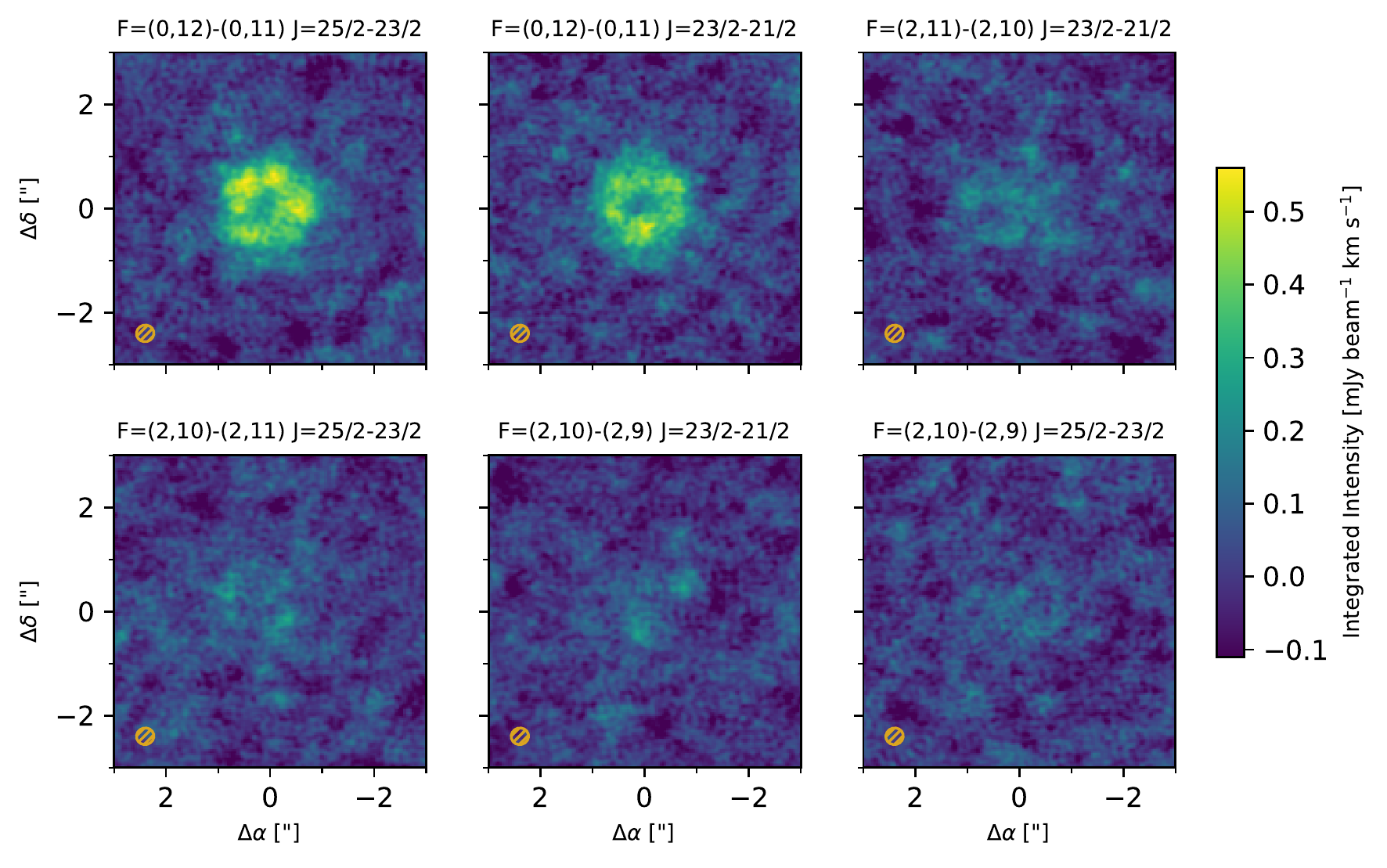}
\caption{Integrated intensity images of individual observed transitions. All the panels share the same intensity scale. The synthesised beam is shown in the bottom left corner. We find $\sigma$ to be $0.06~{\rm mJy~beam}^{-1}~{\rm km~s}^{-1}$.}
\label{fig:channel_maps}
\end{figure*}

\section{Observations}
\label{section: 2}

These observations were taken as part of an ALMA project 2018.A.00021 (PI: Teague), designed to observe $^{12}$CO (2-1), CS (5-4) and CN (2-1) emission at high spatial and spectral resolution from the disk around TW Hya (J2000 R.A. 11h01m51.905s, Decl. -34d42m17.03s). The correlator set-up included a single frequency divided mode (FDM) window centered on 241.5~GHz to provide continuum on which to self-calibrate. In this window six strong emission lines were detected and identified as \ce{CH2CN} lines, and a single \ce{C^34S} line.

\subsection{Data Reduction}

\begin{deluxetable*}{cccccccccc}[hpt!]
\tabletypesize{\footnotesize}
\tablecolumns{10}
\tablecaption{Observed CH$_{2}$CN Transitions
\label{table: Table 1}}
\tablehead{
\colhead{$N^{\prime} - N^{\prime\prime}$}& \colhead{$K_a$} & \colhead{$K_c$} &\colhead{$J^{\prime} - J^{\prime\prime}$} & \colhead{$\nu_0$} & \colhead{$g_u$} & \colhead{$A_{ul}$}  & \colhead{$S_{\rm ij}\mu^2$}  & \colhead{$E_{\rm upper}$}  & \colhead{Int. Flux Density} \\
\colhead{ } & \colhead{ } & & & \colhead{(GHz)} & \colhead{ } & \colhead{$({\rm s}^{-1})$} & \colhead{$({\rm D}^2)$} & (K) & \colhead{$({\rm mJy~km~s}^{-1})$}
}
\startdata
12-11 & 0 - 0 & 12 - 11 & 25/2 - 23/2 & 241.3335423 & 234 & $9.62 \times 10^{-4}$ & 1376.0 & 75  & $126.1 \pm 6.0$ \\ \vspace{0.2cm}
12-11 & 0 - 0 & 12 - 11 & 23/2 - 21/2 & 241.3458390 & 216 & $9.59 \times 10^{-4}$ & 1265.6 & 75  & $109.3 \pm 6.0$ \\ 
12-11 & 2 - 2 & 11 - 10 & 23/2 - 21/2 & 241.3860255 & 216 & $9.33 \times 10^{-4}$ & 1230.5 & 128 & $43.4  \pm 6.0$ \\ \vspace{0.2cm}
12-11 & 2 - 2 & 11 - 10 & 25/2 - 23/2 & 241.3913950 & 234 & $9.36 \times 10^{-4}$ & 1337.7 & 128 & $37.8  \pm 6.0$ \\ 
12-11 & 2 - 2 & 10 - 9  & 23/2 - 21/2 & 241.4925510 & 216 & $9.34 \times 10^{-4}$ & 1230.6 & 128 & $25.2  \pm 6.0$ \tablenotemark{\rm \footnotesize a}\\ 
12-11 & 2 - 2 & 10 - 9  & 25/2 - 23/2 & 241.4970603 & 234 & $9.37 \times 10^{-4}$ & 1337.7 & 128 & $25.2  \pm 6.0$ \\
\enddata
\tablecomments{All data for column $N$' - $N$' through to \textit{$E_{upper}$} was obtained from The Cologne Database for Molecular Spectroscopy \citep[CDMS;][]{Muller_ea_2001}.
\tablenotetext{\rm a}{This peak includes two distinct transitions: the one that we observe and the another one associated with para-\ce{CH2CN}. The latter has a significantly lower $A_{ul}$ ($\sim 3.5\times10^{-6}$s$^{-1}$), thus we assume that all the integrated flux observed comes from the ortho line \citet{Endres_2016}.}
}
\end{deluxetable*}

The data consists of six executions, two in a compact configuration with baselines spanning 15~m -- 500~m on April 4th 2019, and four in a more extended configuration with baselines ranging between 15~m and 2.62~km on September 29th 2019. The shorter baseline executions included 41.7 minutes on-source integration while the longer baseline executions used 51.1 minutes on-source for a total on-source time of 4.9 hours. The quasar J1037-2934 was used for both bandpass and flux calibration for all executions while the phase calibration was performed with J1147-3812 for the short baseline data and J1126-3828 for the long baseline data.

Initial calibration was performed using the standard pipeline procedure in \texttt{CASA} v5.6.2 \citep{McMullin_2007}. The data were then self-calibrated following the self-calibration procedure used in the DSHARP program \citep{Andrews_ea_2018}. In brief, all spectral windows were used, masking out any lines in each spectral window. These line-free observations were used to solve for the phase solutions which were then applied to the entire dataset. Prior to combining the different executions, the executions were aligned to the same phase center and the continuum fluxes were compared. All executions yielded fluxes that were within 2\% of one another, except for the final long baseline execution which varied by about 10\%. The final long baseline execution was rescaled using the \texttt{gaincal} task such that the total flux matched that of the other three long baseline executions. The continuum was subtracted using the \texttt{uvcontsub}.

The continuum FDM window was imaged using the multi-scale CLEAN algorithm and adopting a Briggs weighting with a robust parameter of 2 (similar to natural weighting) yielding a synthesized beam of $0\farcs34 \times 0\farcs32$ at a position angle of 104.8\degr{}. The data was imaged at the native spectral resolution of the FDM window of $1.4~{\rm km\,s^{-1}}$.

Figure \ref{fig:frequency_spectrum} shows the six detected transitions in the form of three sets of doublets with their associated quantum labelling. Since our emission is not spectrally resolved (due to the spectral-set up and face on orientation of the disk), we were unable to extract our spectra using either a matched filtering approach \citep[e.g.,][]{Loomis_ea_2018b} or a line-stacking approach, such as in GoFish \citep{GoFish}. The detected transitions are also depicted as integrated intensity (moment-0) maps in Figure \ref{fig:channel_maps}. While four of the six lines are fairly weak, the two at 241.3335423 and 241.3458390~GHz are robustly detected, exhibiting a clear ring morphology. To better visualize the \ce{CH2CN} morphology, we create a high signal-to-noise map by stacking the 6 transitions together. The resulting image is shown in the left panel of Figure \ref{fig:stacked_radial}. 

For a more direct comparison with the results presented in \citet{Loomis_2018}, we create a version of the images which were smoothed to the same spatial resolution, $1\farcs05 \times 0\farcs83$, using the \texttt{imsmooth} task in \texttt{CASA}. As the weaker transitions are more clearly detected in the smoothed images, we use the smoothed images for both the radially-resolved analysis in \S \ref{section:radial_analysis} and the disk-averaged analysis in \S \ref{section:disk_averaged_analysis}. The higher spatial resolution data is only used in the plotting of the channel maps (Figure \ref{fig:channel_maps}) and of the radial profile (Figure \ref{fig:stacked_radial}) to better constrain the ring morphology. 

\subsection{Observational Results}
\label{section: obs_results}

\begin{figure*}[hpt!]
\centering
\includegraphics[width=0.8\textwidth]{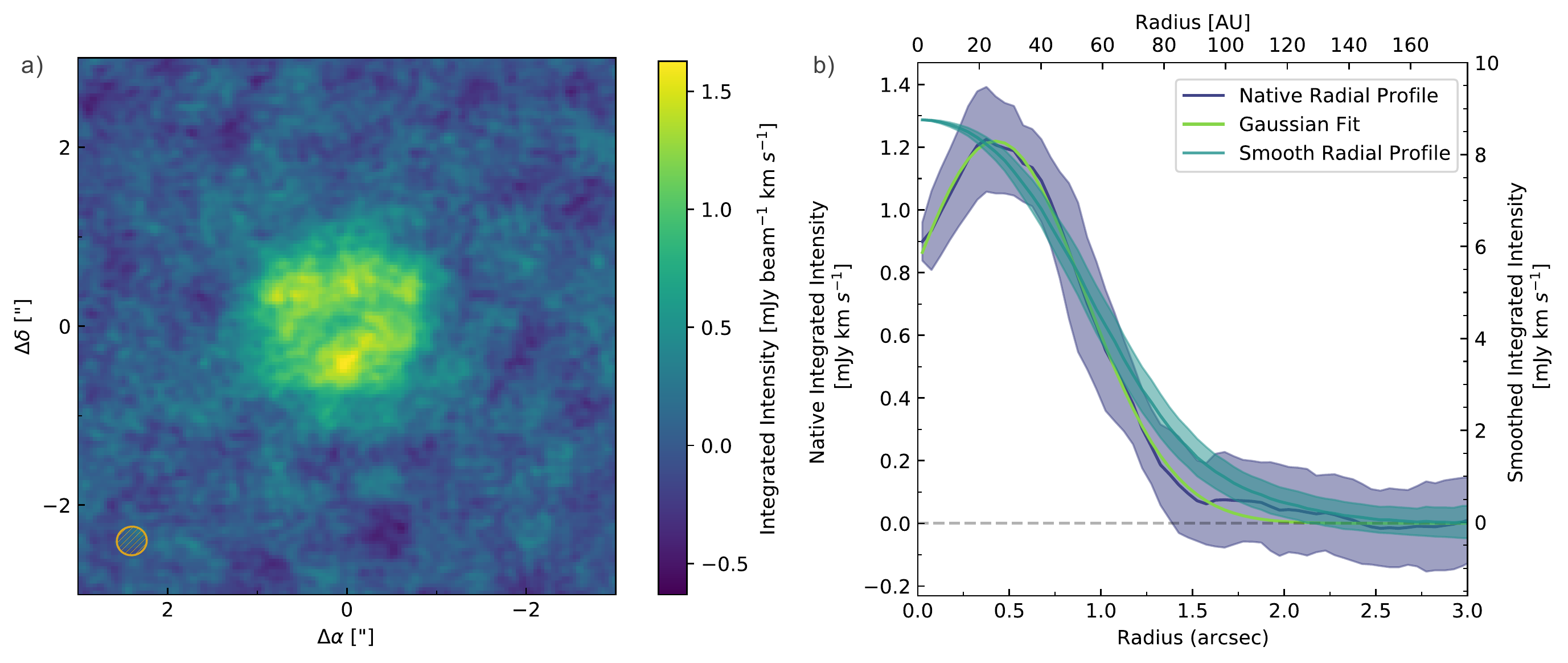}
\caption{Left panel: Integrated intensity map made from the stacked 6 transitions. The noise is calculated to be 0.14 mJy beam$^{-1}$km s$^{-1}$. Right panel: the azimuthally averaged integrated intensity radial profile obtained from the stacked native resolution data. Shaded areas represent $1\sigma$ uncertainties, where $\sigma$ represents the standard deviation in each radial bin.}
\label{fig:stacked_radial}
\end{figure*}

The disk-averaged integrated flux measurements reported in Table \ref{table: Table 1} were obtained by integrating the peaks highlighted in Figure \ref{fig:frequency_spectrum} over the two independent 1.5 km/s channels showing emission out to $\sim 2.5$\arcsec{} (where the intensity reaches 0 in the right panel of Figure \ref{fig:stacked_radial}). Under the assumption of spectrally independent pixels, the uncertainty in the integrated flux is calculated using the equation: 

\begin{equation}
    \delta M_{0} = \sqrt{\sum\limits_{i (I_i > 0)}^N \sigma_{i}^{2} \cdot \Delta v^{2}_{{\rm chan}, i}},
    \label{eqn: spectra_uncertainty}
\end{equation}

\noindent where $\delta M_{0}$ indicates the uncertainty in moment-zero (integrated intensity) values; $\sigma_{i}$ is the signal-to-noise ratio in the spectrum; and $\Delta v_{{\rm chan}, i}$ is the channel width \citep{Teague_2019}. 

We plot radial profiles to see how the integrated flux changes over the radius of the disk. We radially bin the integrated flux from each transition into 0.05\arcsec{} ($\sim$3~au)-wide bins. The beam size was 0.3\arcsec{} ($\sim$18~au), therefore our chosen bin size is a sixth of the beam major FWHM. We use a position angle of 152$\degr$ and an inclination, $i = 5\degr$ \citep{Huang_2018}. We use the native resolution data to plot the radial profile in Figure \ref{fig:stacked_radial}b, which shows the ring morphology quite clearly. For the purpose of our excitation analysis, however, we use the lower spatial resolution smoothed data, with the radial profiles from the individual transitions using this data set shown in Fig. \ref{fig:nt_trot_panel}B.

\noindent The stacked image radial profile shows a ring morphology (Figure \ref{fig:stacked_radial}a), which is also observed in the individual transitions in Fig \ref{fig:channel_maps}. We fitted the radial profile in Figure \ref{fig:stacked_radial} with a Gaussian function (Figure \ref{fig:stacked_radial}b) to infer the location of the center of our ring and the ring width. We find the center to be at 0.4\arcsec{} ($\sim$24~au) and, full width at half maximum of 1.1\arcsec{} ($\sim$69~au). We also see some excess emission between 1.5\arcsec{} and 2.5\arcsec{} when compared to a single Gaussian ring, however, further characterisation of this feature requires more sensitive observations. 

In contrast to the native resolution data, the smoothed data appears to be consistent with a centrally-peaked morphology which is also seen with \ce{CH3CN} \citep{Loomis_2018}. The similar distibution suggests a chemical link between the two molecules, which is explored further in Section \ref{section: discussion}.

\section{\ce{CH2CN} Excitation Analysis}
\label{section: excitation_analysis}

The rotational temperature, $T_{\rm rot}$, and the total column density of \ce{CH2CN}, \textit{N}, can be constrained through the use of a rotational diagram analysis. This implicitly assumes that the molecular excitation can be described by a single temperature and that the molecules are in local thermodynamic equilibrium (LTE). As we assume that the critical densities for \ce{CH2CN} and \ce{CH3CN} are similar ($2.0\times10^6$~cm$^{-3}$ and $2.6\times10^6$~cm$^{-3}$ for the transitions $J = 13-12$ and $J=12-11$ at 40~K, respectively), and that chemical models of TW~Hya suggest that even atmospheric densities exceed $10^9$~cm$^{-3}$ \citep[e.g.,][]{Cleeves_2015}, it is reasonable to assume that both \ce{CH2CN} would be thermalized. As such, we can assume that the excitation temperature is equal to the gas kinetic temperature \citep{Yancy_2015, Guzman_2018, Loomis_2018}. These parameters allow us to draw conclusions about the physical conditions where \ce{CH2CN} is found and its potential interactions with molecules that exist in a similar environment. 

\subsection{Method}
\label{section: method}

\begin{figure*}
\centering
\includegraphics[width=0.9\textwidth]{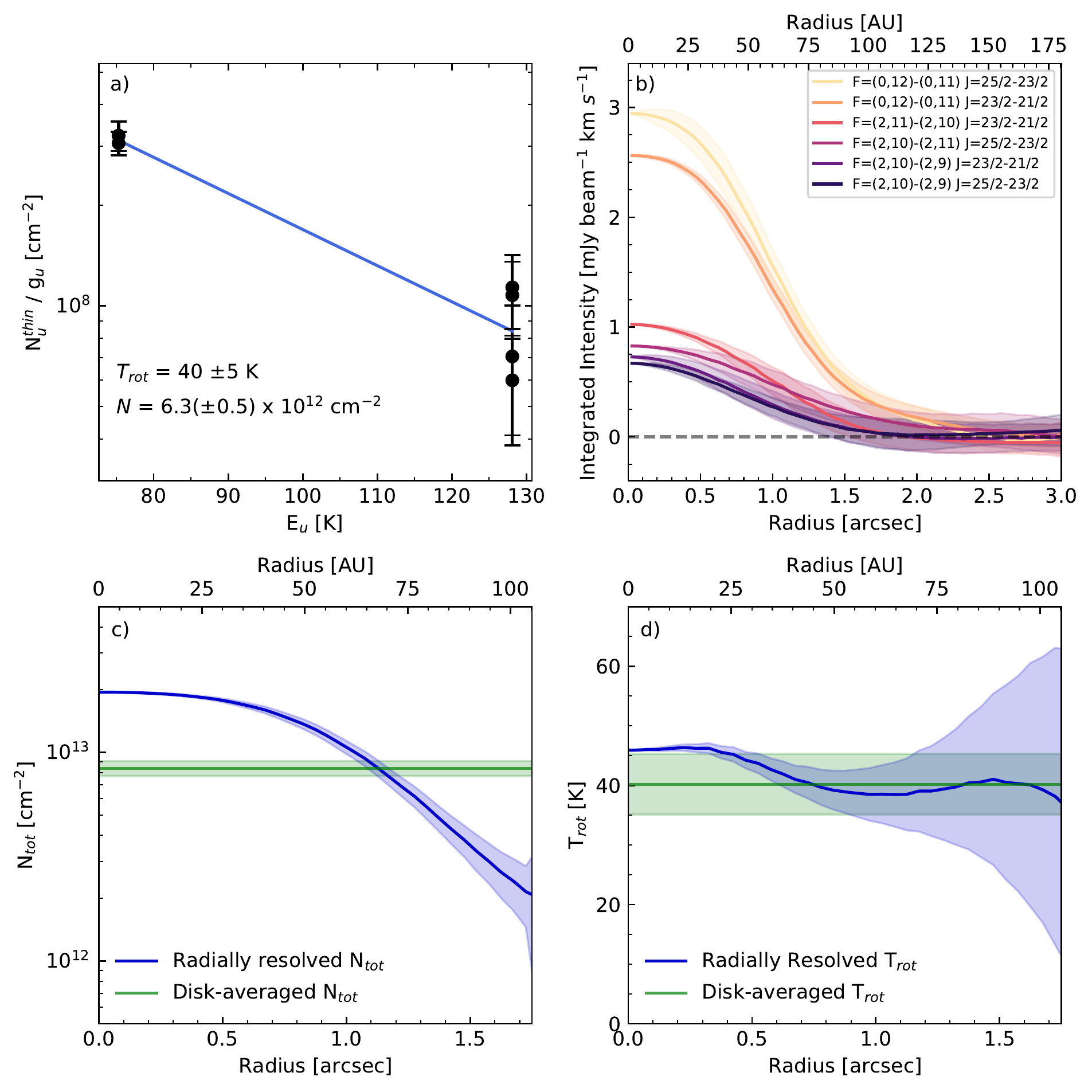}
\caption{Panel A: ortho-\ce{CH2CN} disk-averaged rotational diagram. Panel B: the radial profiles for the six detected transitions. Shaded areas represent 1$\sigma$ uncertainties. Panels C \& D: radial profiles of respectively the \ce{CH2CN} total column density and the rotational temperature. The colour dark blue indicates radially resolved values, whereas green is used for values we obtained from disk-averaged analysis. Shaded areas represent 1$\sigma$ uncertainties.}
\label{fig:nt_trot_panel}
\end{figure*}

We start our analysis by constraining the disk-averaged rotational temperature, $T_{\rm rot}$, and the total column density, $N$. Following \citet{Goldsmith_1999}, we obtain the rotational diagram shown in Figure \ref{fig:nt_trot_panel}A by using the equation

\begin{equation}
    \label{eqn: log_eqn_corrected}
    \ln \frac{{N_{\rm{u}}^{\rm thin}}}{g_{\rm u}} + \ln C_\tau = \ln N - \ln Q(T_{\rm rot}) - \frac{E_{\rm u}}{k T_{\rm rot}}, 
\end{equation}

\noindent where $N_{\rm{u}}^{\rm thin}$ is the column density of molecules in the upper state of each transition without the correction for the optical depth effect, $C_{\tau}$ is the optical correction factor, and $Q(T_{\rm rot})$ is the molecular partition function and $E_{\rm u}$ is the upper state energy. We use the molecular partition function from The Cologne Database for Molecular Spectroscopy \citep[CDMS;][]{Muller_ea_2001} \footnote{Available at https://cdms.astro.uni-koeln.de.} using a linear interpolation to obtain our $Q$ values \citep{Endres_2016}. Degeneracies due to the hyperfine structure are included in the calculation of $Q$. 

We calculate the $N_{\rm{u}}^{\rm thin}$ for each transition through the equation:

\begin{equation}
\label{eqn: N_u}
    N_{\rm{u}}^{\rm thin} = \frac{4 \pi S_v \Delta v}{A_{\rm ul} \Omega h c},
\end{equation}

\noindent where $A_{\rm ul}$ is the Einstein coefficient, $S_\nu$ is the disk-averaged integrated flux density calculated as described in \S \ref{section: obs_results} and using a bin-size of 2.5\arcsec{}, $\Delta$ is the width of the two channels that we used for integration and $\Omega$ is the solid angle subtended by the source.

The optical correction factor is obtained through the equation:

\begin{align}
    \label{eqn:C_tau}
    C_\tau = \frac{\tau}{1 - e^{-\tau}}.
\end{align}

\noindent where the optical depth, $\tau$, can be related to the upper level population through the equation: 

\begin{equation}
    \label{eqn: tau}
    \tau_{\rm ul} = \frac{A_{\rm ul} c^3}{8 \pi \nu^3 \Delta v} N_u (e^{h \nu / k T_{\rm rot}} -1 ).
\end{equation}

Given that our emission is dominated by Doppler broadening, the line width, $\Delta v$, is given by

\begin{equation}
    \label{eqn:thermal_broadening}
    \Delta v = \sqrt{\frac{2 k T_{ \rm rot}}{m_u \, m_H}} ,
\end{equation}

\noindent with $m_{u}$ being the molecular weight of CH$_2$CN (40 g/mol) and $m_H$ being the mass of a hydrogen atom.

\noindent We create a model that relates $N_u$, $C_\tau$ and $\Delta v$ to Equation \ref{eqn: log_eqn_corrected} and we derive the values of $T_{\rm rot}$ and $N$ by matching the observed $N_u/g_u$ values. We use Scipy's curve\_fit function to minimize $\chi^{2}$ and find the best estimate of our desired parameters \citep{Jones_2001}. $\ln(N_u/g_u)$ can then be plotted against the upper state energies, $E_u$ to obtain Figure \ref{fig:nt_trot_panel}A, where the slope and the $y$-intercept of the line respectively represent $-T_{\rm rot}^{-1}$ and $N$.  

\subsection{Disk-averaged Analysis}
\label{section:disk_averaged_analysis}

\begin{deluxetable*}{c|ccccc|c|cccc}[hpt!]
\tablecolumns{11}
\tabletypesize{\footnotesize}
\tablecaption{Formation and destruction pathways for \ce{CH2CN}}
\tablehead{
\multicolumn{1}{c|}{{Reaction}} & \multicolumn{2}{c}{{Reactants}} & & \multicolumn{2}{c|}{{Products}} & \multicolumn{1}{c|}{{Mechanism}} & \multicolumn{4}{c}{Rates of Reaction} \\
\multicolumn{1}{c|}{{Type}} & & & & & & & $\alpha$ $[{\rm cm^3~s^{-1}}]$ & $\beta$ & $\gamma$ & rate type\tablenotemark{\rm \footnotesize c}
}
\startdata
\multirow{6}{*}{Formation}  & \ce{CN}       & \ce{CH3}       & \ce{->}      & \ce{H}        & \ce{CH2CN}    & Neutral-Neutral                           & $1.00(-10)$ & $0.00$      & 0.00 & (1) \\
                            & \ce{N}        & \ce{C2H3}      & \ce{->}      & \ce{H}        & \ce{CH2CN}    & Neutral-Neutral                           & $6.40(-11)$ & $0.17$      & $0.00$ & (1) \\
                            & \ce{C}        & \ce{CH2NH}     & \ce{->}      & \ce{H}        & \ce{CH2CN}    & Neutral-Neutral                           & $1.00(-10)$ & $0.00$      & $0.00$ & (1) \\
                            & \ce{CH3CN+}   & $e^-$          & \ce{->}      & \ce{H}        & \ce{CH2CN}    & DR\tablenotemark{\rm \footnotesize a}     & $2.00(-7)$  & $-0.50$     & $0.00$ & (1) \\
                            & \ce{CH3CNH+}  & $e^-$          & \ce{->}      & \ce{2H}       & \ce{CH2CN}    & DR\tablenotemark{\rm \footnotesize a}     & $8.00(-8)$  & $-0.50$     & $0.00$ & (1) \\
                            & \ce{s-CN}\tablenotemark{\rm \footnotesize b}  & \ce{s-CH2}\tablenotemark{\rm \footnotesize b} & \ce{->} & \multicolumn{2}{c|}{CH$_2$CN} & Grain Surface & & & &  \\\hline
\multirow{5}{*}{Destruction}& \ce{CH2CN}    & \ce{C+}        & \ce{->}      & \ce{C}        & \ce{CH2CN+}   & Ion-Polar                                 & $1.00$      & $1.61(-9)$  & $5.81$ & (2) \\
                            & \ce{CH2CN}    & H$_{3}^{+}$    & \ce{->}      & \ce{H2}       & \ce{CH3CN+}   & Ion-Polar                                 & $1.00$      & $2.94(-9)$  & $5.81$ & (2) \\
                            & \ce{CH2CN}    & h$\nu{}$       & \ce{->}      & \ce{CN}       & \ce{CH2}      & Photodissociation                         & $1.56(-9)$  & $0.00$      & $1.95$ & (3) \\
                            & \ce{CH2CN}    & h$\nu{}$       & \ce{->}      & \ce{CH2CN+}   & \ce{e-}       & Photodissociation                         & $5.29(-10)$ & $0.00$      & $3.11$ & (3) \\
                            & \ce{CH2CN}    &                & \ce{->}      & \multicolumn{2}{c|}{s-CH$_2$CN\tablenotemark{\rm \footnotesize b}} & Freeze-out &     &             &  & \\
\enddata
\tablecomments{Table showing all the possible formation and destruction pathways of \ce{CH2CN} based on the astrochemical disk model by \citet{LeGal_2019}. The rates of reactions are reproduced from the Kinetic Database for Astrochemistry available at \url{http://kida.astrophy.u-bordeaux.fr} \citep{Wakelam_2012}.}
\tablenotetext{\rm a}{DR = Dissociative Recombination}
\tablenotetext{\rm b}{\textit{s-} indicates solid-state reactions, i.e. reactions occurring on the surface of grains.}
\tablenotetext{\rm c}{Rate formulae (1) Modified Arrhenius Equation $k(t) = \alpha(T/300)^{\beta} e^{-\gamma/T}$ (2) Ion-polar rate coefficient computed using Su-Chesnavich capture approach $k(t) = \alpha\beta(0.62+0.4767\gamma(T/300)^{0.5})$ \citep{Woon_Herbst_2009} (3) Photo-dissociation reaction rate $k(t) = \alpha e^{-\gamma A_{v}}$ where $A_v$ is the visual extinction \citep{Draine_1978, vanDishoeck_1994}.}
\label{table:reaction_table}
\end{deluxetable*}

\noindent In this paper, we will denote the column density of ortho-\ce{CH2CN} using \textit{N}, whereas the total column density of \ce{CH2CN} (including both the ortho and the para isomers) will be indicated using $N_{\rm tot}$. We obtain a disk-averaged rotational temperature of $40 \pm 5$~K and a disk-averaged total column density, $N$, of $(6.3\pm0.5)\times10^{12}$~{\rm cm}$^{-2}$ for ortho-\ce{CH2CN}. As described in \citet{LeGal_2017}, for a molecule containing two identical Hydrogen nuclei, such as \ce{CH2CN}, we expect a statistical ortho/para ratio of 3:1, and therefore we infer a total column density, $N_{\rm tot}$, of $(8.4 \pm 0.7) \times 10^{12}$~{\rm cm}$^{-2}$. Indeed, due to the $X^2$B1 symmetry of the ground electronic state of \ce{CH2CN}, the ortho-to-para ratio of \ce{CH2CN} decreases toward the statistical 3:1 value with increasing temperature as \ce{NH2} \citep{LeGal2016}. \ce{CH2CN} is a heavier molecule than \ce{NH2}, therefore, its ortho-to-para ratio will reach the statistical ratio for lower temperatures than \ce{NH2}. Thus, according to Fig.~1 of \cite{LeGal2016}, the relatively high rotational temperature we derived confirms that it is reasonable to consider a 3:1 statistical ortho-para ratio for \ce{CH2CN}. Given this ratio, we can calculate the expected para column density and using this value and the the expected integrated intensity of the para lines that we do not detect (located at 241.353 and 241.381).  

\noindent Finally, having obtained $T_{\rm rot}$, we calculate the optical depth, $\tau$, of our transitions. In all cases, we obtained a value of $\tau \ll 1$, with values of range $4 \times 10^{-4}$ to  $3.4 \times 10^{-3}$. Therefore, our detected transitions were of negligible optical thickness. In addition to this, using these parameters we can calculate the predicted strength of the lines that we do not detect. For the transitions at 241.353 and at 241.382 (which belong to para-\ce{CH2CN}) we obtain integrated intensities of 1.26 and 1.37 mJy km s-1, which are significantly below our calculated intensities, thus confirming that we would not have been able to detect these lines with our spectral set-up. The disk-averaged values obtained in Section \ref{section: method} are calculated out to a radius of 1.75\arcsec{} as we can see from Figure \ref{fig:nt_trot_panel}b that at this radius, all six transitions are detected. As a comparison, we calculate an average $T_{\rm rot}$ and $N$ for the outer region of the disk where the weaker emission lines are almost completely lost. For this, we integrated out to a radius of 2.5\arcsec{}. For this outer region, we find a total column density of $(3.5 \pm 0.3) \times 10^{12}$ ($(4.7 \pm 0.4) \times 10^{12}$ for ortho-\ce{CH2CN}) cm$^{-2}$, and a rotational temperature of $38 \pm 5$~K. Comparing the total column density that we obtained out to a radius 1.75\arcsec{} to that out to 2.5\arcsec{} confirms that the outer radii are contributing very little to the total column density, as we would expect from the radial profiles in Figure \ref{fig:nt_trot_panel}B and in the right panel of Figure \ref{fig:stacked_radial}.

\subsection{Radially Resolved Analysis}
\label{section:radial_analysis}

To further observe the behaviour of $T_{\rm rot}$ and N$_{\rm tot}$ across the disk, we use the radial integrated intensities from each of the transitions in Figure \ref{fig:nt_trot_panel}B and we repeat the steps outlined in Section \ref{section:disk_averaged_analysis} to obtain rotational diagrams at different radii. The results are summarised in panel \ref{fig:nt_trot_panel}C and D. 

Figure \ref{fig:nt_trot_panel}C shows that the column density decreases from $5 \times 10^{13}$ to $0.6 \times 10^{13}$~cm$^{-2}$across the radius of the disk. This is consistent with the disk-averaged column density. The $T_{\rm rot}$ ranges between 45 and 37 K, which is consistent with the disk-averaged $T_{\rm rot}$ (40 $\pm$5 K). We note that this is very similar to the \ce{CH3CN} excitation temperature of $32.7^{+3.9}_{-3.4}$~K in the same disk \citep{Loomis_2018}. On the other hand, the column density of \ce{CH3CN}, $1.82^{+0.25}_{-0.19} \times 10^{12}$~cm$^{-2}$, is $\sim$10 times lower than our observed \ce{CH2CN}'s $N_{\rm tot}$. This is consistent with our chemical model, as discussed in Section \ref{section:chemical_model}.

\section{Discussion}
\label{section: discussion}

\subsection{\ce{CH2CN}/\ce{CH3CN} Ratio \& Disk Models Results}
\label{section:chemical_model}

\begin{figure*}[hpt!]
\label{fig:model}
\centering
\includegraphics[width=0.9\textwidth]{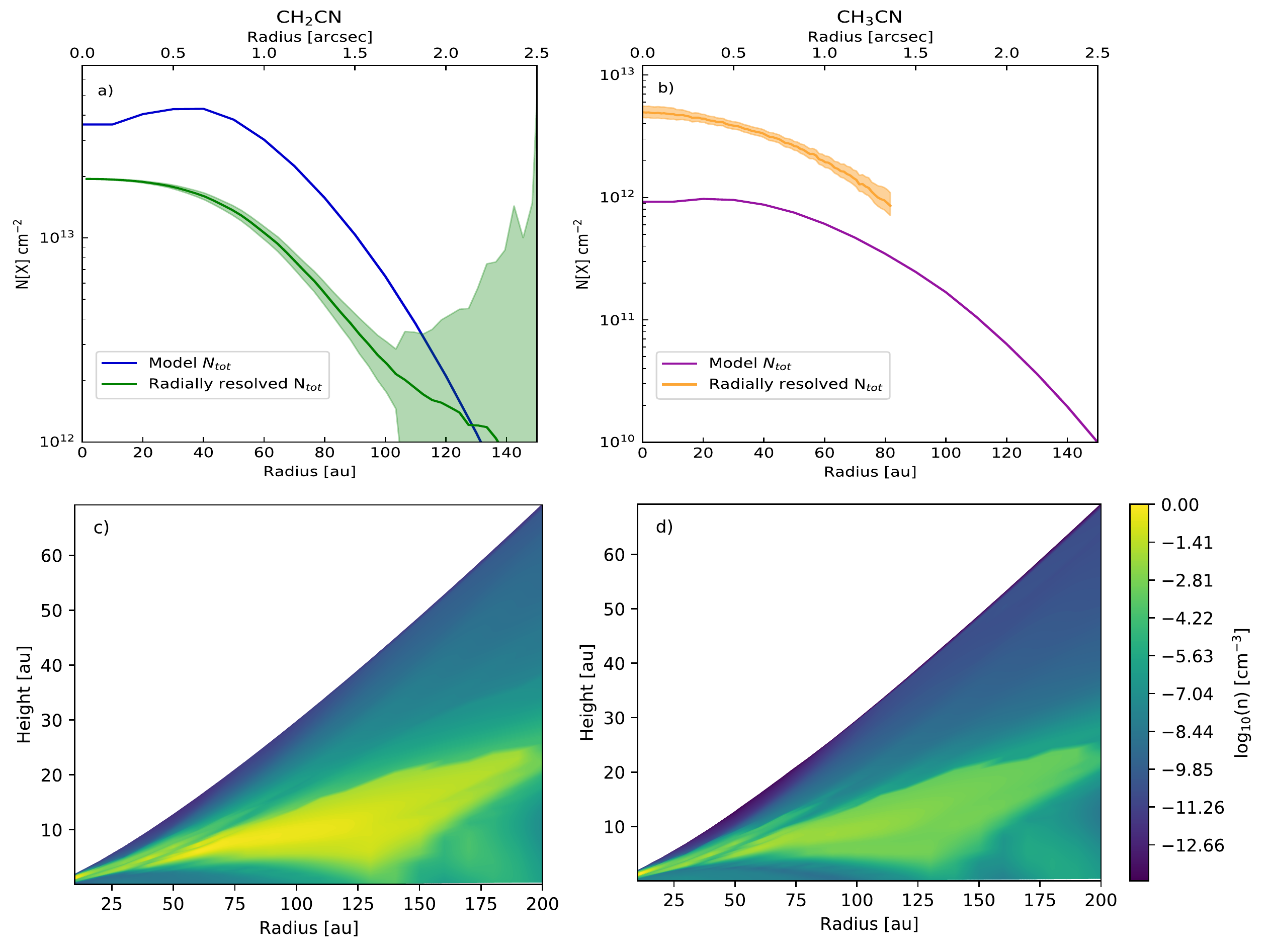}
\caption{Panel a) \& b): Total column densities of \ce{CH2CN} (a) and \ce{CH3CN} (b) as a function of radius, predicted by the chemical models of \citet{LeGal_2019} that we adapted to TW~Hya as described in \S\ref{section:chemical_model}. The column density radial profile derived from the observations presented in Sect.~\ref{section:radial_analysis} for \ce{CH2CN} (a) from the observation of \ce{CH3CN} obtained by \citet{Loomis_2018} (b) are also shown. Panel c) \& d): The abundance of \ce{CH2CN} and \ce{CH3CN} as a function of radius and height. The region of highest emission for both molecules is located between $\sim50-100$~au and at a height of $\sim10$~au. }
\end{figure*}

We use the chemical models by \citet{LeGal_2019} to explore the predicted column densities of \ce{CH2CN} and \ce{CH3CN} as a function of radius, as shown in Figure \ref{fig:model}a \& b. The original model has been modified to fit TW~Hya using the physical parameters described in Table~\ref{tab:mod_TwHya_phys_str}, and a standard cosmic-ray ionization rate of $1\times 10^{-17}$~s$^{-1}$. We used a C/O ratio of 1 which was the best C/O ratio found in \citet{LeGal_2019} to reproduce the column densities of the nitriles detected in abundance in disks. Overall, the model is in good accordance with our observations as it predicts that \ce{CH2CN} is more abundant than \ce{CH3CN} at all radii. However, the \ce{CH2CN}/\ce{CH3CN} ratio in the model is larger than the ratio of $\sim$5 we found observationally. Inspecting Fig. \ref{fig:model}a\&b, we see that the model correctly predicts the column density of \ce{CH2CN} and under-predicts that of \ce{CH3CN}. This suggests that while the column density of \ce{CH2CN} is well reproduced by standard astrochemical model assumptions, there are missing chemical pathways that lead to the formation of \ce{CH3CN}. Detailed astrochemical disk modeling effort needs to be made to further constrain the main reaction pathways driving both the observed \ce{CH3CN} abundance and \ce{CH2CN}/\ce{CH3CN} ratio in the TW~Hya protoplanetary disk, but they are beyond the scope of this paper and will be the subject of a future dedicated study. For now, the use of chemical models allows us to confirm that a ratio of \ce{CH2CN}/\ce{CH3CN}$>$1 is indeed plausible in our disk and allows us to give a more comprehensive overview of \ce{CH2CN} in TW~Hya. Figure \ref{fig:model}c\&d show the abundance distribution of \ce{CH3CN} and \ce{CH2CN} with respect to height and the radius of TW~Hya. Once again, we can see that \ce{CH2CN} is more abundant than \ce{CH3CN} at all radii and heights in the disk and that the region of brightest emission for both of these molecules is co-localised and between a radius of $\sim$50-100~au and a height of $\sim$10~au. This is consistent with our observations as the model in Figure \ref{fig:model}a shows that our observed $N_{\rm tot}$ for \ce{CH2CN} arises from a radius of $\sim$100~au.

\begin{table}
\begin{center}
\caption{Physical parameters used for the disk chemistry modeling  \label{tab:mod_TwHya_phys_str}}
\begin{tabular}{lc}
\hline\hline
Parameters&TW~Hya$^a$\\
\hline
\hline
Stellar mass: $M_\star$ ($M_\odot$) &0.8\\
Characteristic radius: $R_{\rm{c}}$ (AU) &10\\
Density power-law index &1.5\\
Midplane temperature at $R_{\rm{c}}$ (K) &20\\
Atmosphere temperature at $R_{\rm{c}}$ (K) &104\\
Surface density at $R_{\rm{c}}$ (g cm$^{-2}$) &0.79\\
Temperature power-law index &0.55\\
Vertical temperature gradient index &2\\
UV Flux$^b$ at $R_{\rm{c}}$ (in \cite{draine1978}'s units) &3400\\
\hline
\hline
\end{tabular}
\begin{list}{}{}
\item $^a$ \cite{Andrews_2012}
\item $^b$ \cite{Herczeg_2004}
\end{list}
\end{center}
\end{table}

\subsection{\ce{CH2CN} Chemistry}
\label{section:chemistry}

Formation of molecular species in protoplanetary disks can occur via two distinct routes: gas-phase and grain-surface reactions, and in the case of \ce{CH2CN} both pathways are a priori possible. Based on the astrochemical disk model of \citet{LeGal_2019}, several formation and destruction pathways are involved in the chemistry of \ce{CH2CN}. These are summarized in Table \ref{table:reaction_table}. 

As previously mentioned, \ce{CH3CN} and \ce{CH2CN} are found to have similar excitation temperatures and morphologies, therefore it is unsurprising that they also share a comparable chemistry. Both \ce{CH3CN} and \ce{CH2CN} share a very similar grain-surface formation chemistry where both can be formed via solid state reactions with CN as follows,
\begin{align}
    \ce{CH3 + CN & -> CH3CN \\
        CH2 + CN & -> CH2CN \label{eqn:grain_surface}}
\end{align}
\noindent \citep{Herbst_1985}. From these reactions we can infer that the relative formation rates of \ce{CH3CN} and \ce{CH2CN} will depend on the relative abundance of \ce{CH2} and \ce{CH3} radicals on the grain surface. The similarities extend beyond these grain-surface reactions, as these molecules share a common gas-phase reaction pathway. The electronic dissociative recombination of \ce{CH3CNH+} has been identified as the dominant contributor to the gas-phase formation of both \ce{CH3CN} and \ce{CH2CN} \citep{Vastel_2015, Loomis_2018}. The full reaction route, including the formation of the ion intermediate, is as follows:
\begin{align}
    \ce{CH3+ + HCN & -> CH3CNH+ + hv \\
        CH3CNH+ + e- & -> CH2CN + 2H \\
        CH3CNH+ + e- & -> CH3CN + H} 
\end{align}
\noindent \citep{Herbst_1990}. This does not seem to be the main formation pathway for \ce{CH2CN} in either the model or in `reality' since the electronic dissociative recombination (DR) of \ce{CH3CNH+} leads to a branching ratio of $\sim$1.5 in favour of \ce{CH3CN} according to the Kinetic Database for Astrochemistry\footnote{http://kida.astrophy.u-bordeaux.fr} (see Table~\ref{table:reaction_table}) while we find that \ce{CH2CN} $\gg$ \ce{CH3CN}. The branching ratio for this reaction is arbitrary and it is loosely based on the fact that the energy in the \ce{CH3CN + H} channel is greater than the secondary dissociation energy in the \ce{CH2CN + 2H} channel \citep{Wakelam_2012}. In light of this, we suggest that the grain surface formation pathway is the most important source of \ce{CH2CN} in disks. A similar conclusion was made with respect to \ce{CH3CN} by \citep{LeGal_2019} and \citep{Loomis_2018}. However, grain-surface reactions are poorly constrained both theoretically and experimentally and therefore more experimental studies need to be carried out to verify the significance of this solid-state reaction. Finally we note that \ce{CH2CN} can also form through several neutral-neutral reactions (Table \ref{table:reaction_table}), whose possible contributions also warrant further investigation.

\subsection{Morphology}
\label{section:morphology}

\noindent As mentioned in Section \ref{section: obs_results}, \ce{CH2CN} is found in a ring morphology when imaged at the native resolution of $\approx 0.3\arcsec{}$. The ring morphology is not unique to \ce{CH2CN} as CN and \ce{C2H} are also analogously distributed in TW~Hya, however, their rings peak at considerably larger radii at $\sim$ 45~au (0.75\arcsec{}) and 60~au (1\arcsec{}), respectively \citep{Bergin_2016, Teague_2020}. These nested rings of potentially chemically related species provide clues as to what processes are regulating the abundance of these molecules. One possible source of rings is the increased penetration of UV radiation beyond the edge of the dust continuum, leading to more photo-desorption of frozen out molecules from grain surfaces \citep{Cleeves_2016}. However, the edge of the dust continuum is at $\sim 60$~au, which is inconsistent with the peak of the \ce{CH2CN} ring ($\sim 24$~au) therefore this is not a plausible explanation for the morphology of \ce{CH2CN}.

\noindent The chemical model results for \ce{CH2CN} shown in Figure \ref{fig:model} seem to replicate the morphology that we observed in Figure \ref{fig:nt_trot_panel}C. This suggests the presence of a `Goldilocks zone' for the dominant formation pathway where the observed morphology can be attributed to the balance between formation and destruction reactions under fiducial disk conditions. Within this zone of the disk, the conditions may be particularly favourable for the production of \ce{CH2CN} (i.e. optimal UV radiation flux or temperature of the disk molecular layer).

\noindent Another possible factor that could shape the distribution of \ce{CH2CN} is enhanced destruction in the inner disk due to reactions with gas-phase carbon and oxygen atoms. Chemical modeling by \citet{Du_2015} has shown that C and O depletion is needed to reproduce the observed molecular abundances of CO and \ce{H2O} in TW~Hya. Further, nitrile column density is enhanced by $\sim$2 orders of magnitude where C and O depletion takes place \citep{Du_2015}. One disk location that may present a rapid change in the C and O abundance is the CO snowline, where CO ice sublimates back into the gas-phase. The CO snowline zone in TW~Hya is found at 17-30~au \citep{Schwarz_2016, Qi_2013}. We find that \ce{CH2CN} peaks at $\sim$ 24~au, so the CO snowline and the \ce{CH2CN} peak may reflect the presence of C- and O-depleted gas just exterior to the CO snowline. Exterior to this region, carbon and oxygen would primarily exist as CO ice, therefore not interfere with nitrile chemistry. 

\section{Summary}
\label{section: summary}
We have presented the first detection of \ce{CH2CN} (cyanomethyl) in a protoplanetary disk. The 6 emission lines detected correspond to 6 transitions of \ce{CH2CN} in its ortho state. We find a disk-averaged rotational temperature of $40 \pm 5$~K and a disk-averaged total column density of $(6.3\pm0.5 \times 10^{12}$~cm$^{-2}$. Assuming a thermal ortho/para ratio of 3:1, we infer a total \ce{CH2CN} column density of $(8.4\pm0.7)\times 10^{12}$ cm$^{-2}$. The molecule is observed to be in a ring whereas at lower spatial resolutions it presents a centrally-peaked profile consistent with previously reported \ce{CH3CN} morphology.  Comparison with \ce{CH3CN} total column density shows that \ce{CH2CN} is 10 times more abundant.  This result is consistent with chemical models for TW Hya, where \ce{CH2CN}$\gg$\ce{CH3CN} at all disk radii. We identify possible pathways that contribute to the formation
and destruction of this molecule and we suggest that grain-surface reactions are the likely formation pathway for both \ce{CH2CN} and \ce{CH3CN}.

\acknowledgements
We thank the anonymous referees for their valuable input in clarifying the results presented in this paper. We also wish to thank Ryan Loomis for sharing radial profiles of \ce{CH3CN}. This paper makes use of the following ALMA data: ADS/JAO.ALMA\#2018.A.00021. ALMA is a partnership of ESO (representing its member states), NSF (USA) and NINS (Japan), together with NRC (Canada), MOST and ASIAA (Taiwan), and KASI (Republic of Korea), in cooperation with the Republic of Chile. The Joint ALMA Observatory is operated by ESO, AUI/NRAO and NAOJ. The National Radio Astronomy Observatory is a facility of the National Science
 Foundation operated under cooperative agreement by Associated Universities, Inc. AC acknowledges funding from the Origins of Life Initiative. This work was supported by an award from the Simons Foundation (SCOL \# 321183, K\"O).

\bibliography{bibliography}
\bibliographystyle{aasjournal}

\end{document}